%% file: ijcai24.tex
\documentclass{article}
\usepackage{ijcai24}

\usepackage{times}
\usepackage{soul}
\usepackage{url}
\usepackage[hidelinks]{hyperref}
\usepackage[utf8]{inputenc}
\usepackage[small]{caption}
\usepackage{graphicx}
\usepackage{amsmath}
\usepackage{amsthm}
\usepackage{booktabs}
\usepackage{algorithm}
\usepackage{algpseudocode}
\usepackage{float}
\usepackage{svg}
\usepackage[switch]{lineno}

\algnewcommand\algorithmicinput{\textbf{Input:}}
\algnewcommand\Input{\item[\algorithmicinput]}

\usepackage{subcaption}
\usepackage{tikz}
\usepackage{enumitem}
\usepackage{mathtools}
\usepackage{cleveref}
\usepackage{amsfonts}

\usetikzlibrary{shapes.geometric}

\newtheorem{theorem}{Theorem}
\newtheorem{definition}{Definition}

\title{Look-ahead Search on Top of Policy Networks in Imperfect Information Games}

\author{
Ondřej Kubíček$^1$
\and
Neil Burch$^{2,3}$\And
Viliam Lis\'{y}$^{1}$
\affiliations
$^1$Artificial Intelligence Center, Department of Computer Science, Faculty of Electrical Engineering, Czech Technical University in Prague\\
$^2$Sony AI\\
$^3$Alberta Machine Intelligence Institute,   University of Alberta
\emails
\{kubicon3, viliam.lisy\}@fel.cvut.cz,
nburch@ualberta.ca
}
\newcommand{\AlgorithmName}{Multi-Agent Search With Policy Transformations}
\newcommand{\AlgorithmShort}{SePoT}

\newcommand{\FOGGame}{\mathcal{G}}
\newcommand{\GadgetGame}{\mathcal{G}^G}
\newcommand{\Player}{i}
\newcommand{\Opponent}{o}
\newcommand{\Players}{\mathcal{N}}
\newcommand{\WorldState}{w}
\newcommand{\WorldStates}{\mathcal{W}}
\newcommand{\Playerfunction}{p}
\newcommand{\InitState}{w^0}
\newcommand{\ArtificalState}{w^A}
\newcommand{\Action}{a}
\newcommand{\Actions}{\mathcal{A}}
\newcommand{\ActionDistribution}{\Delta \mathcal{A}}
\newcommand{\Transitions}{\mathcal{T}}
\newcommand{\Rewards}{\mathcal{R}}
\newcommand{\ExpectedUtilityFOSG}{u}
\newcommand{\CounterfactualValues}{v}
\newcommand{\Reward}{R}
\newcommand{\ObservationSet}{\mathbb{O}}
\newcommand{\Observations}{\mathcal{O}}
\newcommand{\History}{h}
\newcommand{\InitHistory}{h^0}
\newcommand{\Histories}{\mathcal{H}}
\newcommand{\Terminals}{\mathcal{Z}}

\newcommand{\RealNumbers}{\mathbb{R}}
\newcommand{\PublicIndex}{0}
\newcommand{\ChancePlayer}{c}
\newcommand{\Trajectory}{\tau}
\newcommand{\Isets}{\mathcal{S}}
\newcommand{\Iset}{s}
\newcommand{\Policy}{\pi}
\newcommand{\Policies}{\Pi}

\newcommand{\PolicyTransformation}{f^T}
\newcommand{\PolicyTransformations}{F^T}

\newcommand{\PlayerReach}{P}
\newcommand{\Range}{r}
\newcommand{\IsetDistribution}{\Delta \mathcal{S}}
\newcommand{\PublicBeliefState}{\beta}
\newcommand{\Regularized}{R}
\newcommand{\RegularizationTerm}{\eta}

\newcommand{\SamplingPolicy}{\mu}

\newcommand{\timestep}{t}
\newcommand{\smallertimestep}{q}

\newcommand{\VtraceVal}{v}

\newcommand{\ExpectedValue}{\mathbb{E}}

\newcommand{\VtraceOperator}{\mathcal{V}}
\newcommand{\DiscountFactor}{}
\newcommand{\ContractionConstant}{\lambda}
\newcommand{\TrajectoryLength}{l}

\newcommand{\ContractionProofConstant}{\kappa}

\newcommand{\Exploitability}{\mathcal{E}}
\newcommand{\BestResponse}{{BR}}
\newcommand{\NashEquilibrium}{{NE}}

\newcommand{\ValueFunction}{\boldsymbol{v}}
\newcommand{\OptimalValueFunction}{\ValueFunction^*}
\newcommand{\CriticNetwork}{\boldsymbol{u}}

\newcommand{\PseudocodeTrajectory}{\tau}
\newcommand{\PseudocodeSamplingPolicy}{\mu}
\newcommand{\PseudocodePolicyWeights}{\theta^\pi}
\newcommand{\PseudocodeTransformationWeights}{\theta^t}
\newcommand{\PseudocodeCriticWeights}{\theta^v}
\newcommand{\PseudocodePolicy}{\Policy}
\newcommand{\PseudocodePolicyTransformed}{\Policy^t}
\newcommand{\PseudocodePolicyDirection}{d}
\newcommand{\PseudocodeIterations}{T}
\newcommand{\PseudocodeIset}{\Iset}
\newcommand{\PseudocodePlayer}{\Player}
\newcommand{\PseudocodeRange}{\Range}
\newcommand{\PseudocodeCriticValues}{u}
\newcommand{\PseudocodeCounterfactualValues}{v}
\newcommand{\PseudocodeGadgetGame}{\GadgetGame}

\begin{document}

\include{main_part}

\section*{Acknowledgments}
This research is supported by Czech Science Foundation (GA22-26655S) and Grant Agency of the CTU in Prague (SGS22/168/OHK3/3T/13), and the Canadian Institute for Advanced Research. Computational resources were supplied by (e-INFRA CZ LM2018140) supported by the Ministry of Education, Youth and Sports of the Czech Republic and also (CZ.02.1.01/0.0/0.0/16 019/0000765).

\bibliographystyle{named}
\bibliography{references}

% \appendix
\include{appendix}


\end{document}

%% file: main_part.tex
\maketitle
% \hspace{-100pt}
% \vspace*{-2cm}

\begin{abstract}
Search in test time is often used to improve the performance of reinforcement learning algorithms. Performing theoretically sound search in fully adversarial two-player games with imperfect information is notoriously difficult and requires a complicated training process. We present a method for adding test-time search to an arbitrary policy-gradient algorithm that learns from sampled trajectories. Besides the policy network, the algorithm trains an additional critic network, which estimates the expected values of players following various transformations of the policies given by the policy network. These values are then used for depth-limited search. We show how the values from this critic can create a value function for imperfect information games. Moreover, they can be used to compute the summary statistics necessary to start the search from an arbitrary decision point in the game. The presented algorithm is scalable to very large games since it does not require any search during train time. We evaluate the algorithm's performance when trained along Regularized Nash Dynamics, and we evaluate the benefit of using the search in the standard benchmark game of Leduc hold'em, multiple variants of imperfect information Goofspiel, and Battleships.
\end{abstract}

\section{Introduction} \label{s:intro}
The field of multi-agent deep reinforcement learning has achieved remarkable success in training strong agents for two-player adversarial games through selfplay. Many of these agents develop strategies represented by neural networks, which subsequently provide probability distributions for actions based on current observations during gameplay.

% Despite their success, the neural networks can occasionally lead the agent to make suboptimal decisions that an adversarial opponent may exploit. \cite{katagoadverse} have illustrated how policy networks trained via AlphaZero-style algorithms in Go are susceptible to exploitation, as evidenced by winning against the trained policy network 97\% of the time. Similarly, \cite{perolat2022mastering} report that their trained agents display evident gameplay errors, necessitating the application of multiple heuristics to rectify these errors. However, there are principled ways of using search to avoid these mistakes, as evidenced by AlphaZero \cite{alphazero}, DeepStack \cite{deepstack}, and Student of Games \cite{pog}.

Despite their success, the neural networks can occasionally lead the agent to make suboptimal decisions that an adversarial opponent may exploit. AlphaZero-style algorithms in Go are susceptible to exploitation, as evidenced by winning against the trained policy network 97\% of the time \cite{katagoadverse}. Similarly, policy-gradient algorithms display evident gameplay errors, necessitating the application of multiple heuristics to rectify these errors \cite{perolat2022mastering}. However, there are principled ways of using search to avoid these mistakes, as evidenced by AlphaZero \cite{alphazero}, DeepStack \cite{deepstack}, and Student of Games \cite{pog}.

Algorithms that use policy-gradient methods in imperfect information games like Regularized Nash Dynamics (RNaD) \cite{perolat2022mastering,poincarre} rely on directly applying the neural network's policy during gameplay. Adding search to these methods could improve the strategy and mitigate some mistakes. However, this enhancement comes at the expense of turning model-free algorithms into model-dependent ones during gameplay.

Adding search to policy-gradient methods in imperfect information games presents multiple challenges that are not present in a perfect information case, which makes algorithms like AlphaZero unusable in this setting. First, the imperfect information search must be performed not only from the current unobserved state of the game, but from all possible states that share the same public information amongst all players. Second, the state values depend on beliefs players have about the current information of their opponents. Thirdly, the opponent's strategy is unknown, and the optimal strategies are not unique. Hence, we need to optimize over all possible strategies the opponent may deploy both in the remainder of the game as well as unobserved past actions.

In this paper, we introduce the algorithm \AlgorithmName{} (\AlgorithmShort), which builds upon the concept of multi-valued states \cite{MVS}. \AlgorithmShort{} trains a policy transformation critic network that estimates the expected value of policies derived from the policy network. This critic is trained alongside a policy network with a policy-gradient technique without requiring any additional trajectories beyond those sampled by the original policy-gradient algorithm. The critic network is trained with an off-policy V-trace that estimates expected values against a multitude of potential opponent strategies derived from the policy represented by the policy network. We show that this critic provides sufficient information to perform a safe search \cite{soundsearch} in imperfect information games.

Furthermore, our approach enables safe search at any point within a game,  utilizing the information provided by the introduced critic network. This eliminates the need for conducting search in all decision points from the root until the current decision point, which is often intractable in games with little public knowledge, but at the cost of requiring a domain-specific function that reconstructs current game segment. It is in strict contrast to prior methods \cite{MVS,deepstack,pog}. This flexibility allows us to perform search only in game segments sufficiently simple for available compute power, while employing the neural network policy in other parts of the game.
 
Finally, we establish a connection between value functions and multi-valued states within imperfect information games that enables training previously used imperfect information value functions \cite{deepstack,pog} without the expensive search in the training time.

In our experiments, we evaluate \AlgorithmShort{} with RNaD \cite{perolat2022mastering}. We show that using search in head-to-head play in imperfect-information Goofspiel and Battleships improves the performance over the network trained with RNaD. Furthermore, we demonstrate that the exploitability of policy derived from the search is the same or lower than that of the policy network from RNaD in a small version of imperfect information Goofspiel and the standard benchmark game Leduc hold'em.

Our main contributions are: (1) Adding safe search on top of the policy-gradient method without requiring training time search or additional training trajectories. (2) Introducing search technique that may be performed anywhere in the game without performing search in previous parts of the game. (3) Establishing a practical connection between the multi-valued states and the value functions. (4) Showing that the additional search improves the performance of RNaD.
\section{Issues With Imperfect Information Search} \label{s:motivation}
\begin{figure}[h]
  \centering
  \begin{subfigure}{.12\textwidth}
      \centering
      \includegraphics[width=.8\linewidth]{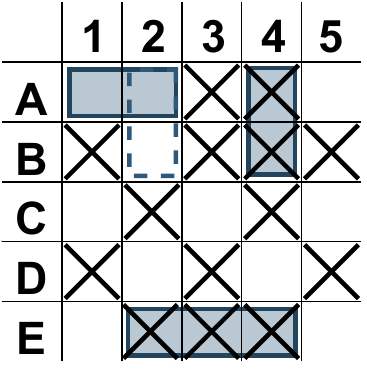}
      \caption{Player}
      \label{fig:bsexampleboard1}
  \end{subfigure}
  \begin{subfigure}{.12\textwidth}
      \centering
      \includegraphics[width=.8\linewidth]{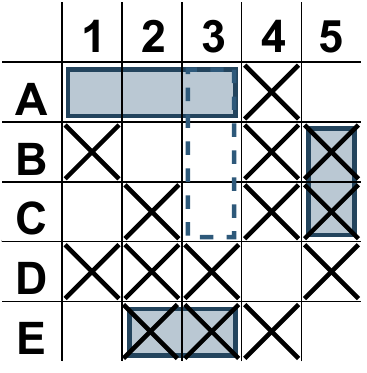}
      \caption{Opponent}
      \label{fig:bsexampleboard2}
  \end{subfigure}
  \begin{subfigure}{.17\textwidth}
  \centering
      \begin{center}
        \resizebox{2.6cm}{!}{
        % \centering
        \begin{tabular}{c|c|c}
        State & P1 ship & P2 ship \\ \hline
        $w_1$  & A1-A2     & A1-A3     \\
        $w_2$  & A1-A2     & A3-C3     \\
        $w_3$  & A2-B2     & A1-A3     \\
        $w_4$  & A2-B2     & A3-C3    
        \end{tabular}
        }
        \end{center}
      \caption{Possible States}
      \label{fig:bsexamplestates}
  \end{subfigure}
  \caption{Current state in a smaller version of Battleships and all possible states sharing the same public information}
  \label{fig:bsexample}
  % \Description{Board state in smaller version of Battleships,}
\end{figure}
Consider a smaller version of Battleships, in which two players have three ships of sizes 3, 2, and 2 on a square grid of size 5. Players take turns shooting on a tile of the opponent's board after all ships are secretly placed. The player that first hits all tiles occupied by the opponent's ships wins. Suppose we are in the middle of the game, and the current boards are shown in \Cref{fig:bsexample}. Each player knows their remaining ship's position, but does not know the opponent's. Based on the public shooting actions up until this point, we know that there are only two possible positions for the player and the opponent. The player may have either placed the ship at A1-A2 or A2-B2. Similarly, the opponent may have placed the ship at A1-A3 or A3-C3. Possible positions are shown on each board with the dashed rectangle. There are four possible states, that is, all possible combinations between the positions across both players, as shown in \Cref{fig:bsexamplestates}. The first player cannot distinguish between states $\WorldState_1$ and $\WorldState_2$, neither between $\WorldState_3, \WorldState_4$. Similarly, the opponent cannot distinguish between $\WorldState_1, \WorldState_3$ and between $\WorldState_2, \WorldState_4$. Shooting anywhere else than A1, A2, A3, B3, and C3 does not make sense for a rational player. However, training RNaD for three days results in a policy that shoots to these tiles with a probability less than $45\%$.

Employing additional search should improve this policy, but doing so naively in imperfect information games introduces multiple issues. The first issue is that starting search only from states that are possible from the player's point of view, which are $\WorldState_1$ and $\WorldState_2$, would search only in those parts of the game where the player placed the ship at A1-A2. This means that the player incorrectly assumes the opponent knows the ship's position during this search. The outcome of the search would be that the game always ends in 2 turns because the opponent would shoot at the ship. Since the player requires at least three turns to win, it could never win against this informed opponent, and its strategy would be to play randomly. However, in the real gameplay scenario, the opponent lacks knowledge of the player's ship's location, providing the player, who is moving first, with a genuine opportunity to win. Nevertheless, this issue can be fixed by starting the search from all four possible states. 

The second issue is more profound. In practice, for many different reasons, the player may incorrectly assume that the opponent is slightly more likely to place his ship at position A1 instead of C3. In such a case, the player would always shoot at either tile A1, A2, or A3 since it provides more reward in expectation from its point of view. This leads to a strategy that could be exploited by an adversarial opponent who would never put the ship there in the first place. This is a well-studied problem in imperfect information games. One possible solution during search is to assume that the opponent can play any strategy in earlier parts of the game. This assumption then allows computing a more robust strategy against possible play from the game's previous and subsequent parts \cite{cfrd,maxmargin,reachmaxmargin}. 

The third issue for applying search in the situation in \Cref{fig:bsexample} are sizes of required game structures. The full end-game cannot be solved online in reasonable time since it has over $10^9$ terminal histories. Also, it is impossible to safely search in earlier decision points, since after the first move there are more than $10^8$ possible ship configurations and branching factor of $50$, which means that even a depth-limited game with look-ahead of one move has more than $10^{10}$ histories. Using a depth-limited search with a value function at the depth limit only later in the game enables search even with much less computational resources \cite{valuefunctions}.
\section{Background} \label{s:background}

Factored-observation stochastic game (FOSG) \cite{FOSG} is a tuple $\FOGGame = (\Players, \WorldStates, \Playerfunction, \InitState, \Actions, \Transitions, \Rewards, \Observations)$, where $\Players = \{1, \dots, N\}$ is a player set, $\WorldStates$ is the set of states and $\InitState$ is an initial state, $\Playerfunction: \WorldStates \rightarrow 2^{\Players}$ is a player function, which for given state $\WorldState \in \WorldStates$ assigns currently acting players from $\Players$, $\Actions = \Pi_{\Player \in \Players} \Actions_\Player$ is the space of joint actions. We denote $\Actions_\Player(\WorldState)$ legal actions for player $\Player$ in the world state $\WorldState$. $\Transitions: \WorldStates \times \Actions \rightarrow \WorldStates$ is the transition function, $\Rewards: \WorldStates \times \Actions \rightarrow \RealNumbers^\Players$ is the reward function and $\Observations: \WorldStates \times \Actions \times \WorldStates \rightarrow \ObservationSet$ is the observation function. $\Observations$ is factored as $\Observations = (\Observations_\PublicIndex, \Observations_1, \dots \Observations_N)$, where $\Observations_\PublicIndex$ are a public observations given to all players and $\Observations_\Player$ are a private observations of player $\Player$. In this work, we focus on two-player zero-sum games, where $\Players = \{1, 2, \ChancePlayer\}$ and $\Rewards_1(\WorldState, \Action) = -\Rewards_2(\WorldState, \Action) \; \forall (\WorldState, \Action) \in \WorldStates \times \Actions$. $\ChancePlayer$ denotes a chance player who has publicly announced a non-changeable policy throughout the game. Moreover, we study only games with perfect recall, where neither player forgets any information.

Trajectory $\Trajectory = \WorldState^0\Action^0\WorldState^1\Action^1 \dots \WorldState^t \in (\WorldStates\Actions)^*\WorldStates$ for which $\Action^\TrajectoryLength \in \Actions(\WorldState^\TrajectoryLength)$ and $\WorldState^{\TrajectoryLength+1} = \Transitions(\WorldState^\TrajectoryLength, \Action^\TrajectoryLength)$ is a finite sequence of states and actions in the game. Trajectory utility for player $\Player$ is a cumulative reward $\ExpectedUtilityFOSG_\Player(\Trajectory) = \sum_{\timestep = 0}^{\TrajectoryLength - 1} \Rewards_\Player(\WorldState^\timestep, \Action^\timestep)$. We use $\Histories$ to denote histories, which are trajectories that start in the initial state $\InitState$. $\Terminals = \{z \in \Histories | z \text{ is terminal} \}$ are histories in which neither player has an action to play, effectively ending the game. $\History \sqsubseteq \History'$ means that $\History'$ extends $\History$ and $\InitHistory$ is an initial history, that corresponds to the initial state $\InitState$, before any player played any action. Each history ends with some world state $\WorldState^\TrajectoryLength$. Therefore, we denote $\Rewards_\Player(\History, \Action) = \Rewards_\Player(\WorldState^\TrajectoryLength, \Action)$ as a reward given to player $\Player$, if in history $\History \in \Histories$ the players play joint action $\Action$. Since player $\Player$ does not observe the state of the game directly, but only observes its actions $\Actions_\Player$, private observations $\Observations_\Player$ and public observations $\Observations_\PublicIndex$, it cannot distinguish between multiple different histories. We say that these indistinguishable histories are in the same infoset $\Iset_\Player \in \Isets_\Player$. If two histories $\History, \History' \in \Histories$ belong to the same infoset $\Iset_\Player$, then player $\Player$ has to have the same actions in these histories and $\Iset_\Player(\History)$ denotes an infoset of player $\Player$ in history $\History$. We will use $\Actions_\Player(\Iset_\Player)$ to denote these available actions in infoset $\Iset_\Player$. Furthermore, we will use a notion of public-states $\Iset_\PublicIndex \in \Isets_\PublicIndex$, which may be viewed as an information set for an external player which does not act in the game. Two histories belong to the same public state if and only if they have the same sequence of public observations from the initial state. Each public state is made from multiple information sets for a given player $\Player$. We will use $\Isets_\Player(\Iset_\PublicIndex)$ to denote all infosets that are in public state $\Iset_\PublicIndex$ and $\Histories(\Iset)$ to denote all histories in an infoset $\Iset$. 

Policy of player $\Player$ is a mapping $\Policy_\Player : \Isets_\Player \rightarrow \ActionDistribution_\Player$ s.t. $\Policy_\Player(\Iset_\Player) \in \ActionDistribution_\Player(\Iset_\Player)$. $\Policies_\Player$ is a set of all policies of player $\Player$. We will use $\Policy_\Player(\Iset_\Player, \Action_\Player)$ to denote the probability of playing action $\Action_\Player$ in $\Iset_\Player$, while following the policy $\Policy_\Player$. A strategy profile is defined as $\Policy = (\Policy_1, \dots, \Policy_N)$, and $\Policy_{-\Player}$ is used to denote a strategy profile without the policy of player $\Player$.

Expected utility, if all players play according to strategy profile $\Policy$ from history $\History$ onward is defined as $\ExpectedUtilityFOSG^\Policy_\Player(\History) = \ExpectedValue_{\Trajectory \sim \Policy | \History}\ExpectedUtilityFOSG_\Player(\Trajectory)$. $\PlayerReach^\Policy(\History) = \prod_{\History'\Action\WorldState \sqsubseteq \History} \prod_{j \in \Players} \Policy_j(\Iset_j(\History'), \Action_j)$ is a probability of reaching history $\History$ if all players play according to strategy profile $\Policy$. This may be separated into $\PlayerReach^\Policy(\History) = \PlayerReach^\Policy_\Player(\History) \PlayerReach^\Policy_{-\Player}(\History)$, where $\PlayerReach^\Policy_\Player(\History)$ is only a players $\Player$ contribution to reaching history $\History$, similarly $\PlayerReach^\Policy_{-\Player}(\History)$ is a contribution of all the other players except $\Player$, which is called counterfactual reach probability. Counterfactual values for infoset $\Iset_\Player$ is a sum of expected utilities of each history $\History$ within the same infoset, weighted by the counterfactual reach probability of this history $\CounterfactualValues_{\Player}^\Policy(\Iset_\Player) = \frac{\sum_{\History \in \Histories(\Iset_\Player)} \PlayerReach^\Policy_{-\Player}(\History)\ExpectedUtilityFOSG_\Player^\Policy(\History)}{\sum_{\History \in \Histories(\Iset_\Player)} \PlayerReach^\Policy_{-\Player}(\History)} $. 
%$\CounterfactualValues_{\Player}^\Policy(\Iset_\Player) = \frac{\sum_{\Iset_\Player(\History) = \Iset_\Player} \PlayerReach^\Policy_{-\Player}(\History)\ExpectedUtilityFOSG_\Player^\Policy(\History)}{\sum_{\Iset_\Player(\History) = \Iset_\Player} \PlayerReach^\Policy_{-\Player}(\History)} $. 

Best response against policy $\Policy_{\Player}$ of a player $\Player$ is a policy $\Policy_{-\Player}^{\BestResponse} \in \BestResponse_\Player(\Policy_{\Player})$, where $\ExpectedUtilityFOSG^{(\Policy_\Player, \Policy^{\BestResponse}_{-\Player})}_\Player(\InitHistory) \geq \ExpectedUtilityFOSG^{(\Policy_\Player, \Policy_{-\Player}')}_\Player(\InitHistory)$ for all $\Policy_{-\Player}'$. If all players play a best response to each other, then the resulting policy profile $\Policy^{\NashEquilibrium}$ is called Nash equilibrium. The usual metric to compare strategies in two-player zero-sum games is an exploitability $\Exploitability(\Policy_\Player) = \ExpectedUtilityFOSG^{(\Policy_\Player, \Policy^\BestResponse_{-\Player})}_{-\Player}(\InitHistory) - \ExpectedUtilityFOSG^{\Policy^\NashEquilibrium}_{-\Player}(\InitHistory)$, that is how much utility can opponent get if it plays best response to the player's strategy, compared to the utility of Nash equilibrium. From definition, if player $\Player$ plays Nash equilibrium, then the exploitability is 0.

\subsection{Depth-Limited Solving} \label{s:dlsolving}
Range $\Range_\Player \in \IsetDistribution_\Player(\Iset_\PublicIndex)$ for player $\Player$ in public state $\Iset_\PublicIndex$ is a probability distribution over infosets $\Iset_\Player~\in~\Isets_\Player(\Iset_\PublicIndex)$ in public state $\Iset_\PublicIndex$, which corresponds to player's contribution to the probability of reaching this infoset. Furthermore range for public state $\Iset_\PublicIndex$ is defined as $\Range = (\Range_1, \dots, \Range_N) \in \IsetDistribution_1(\Iset_\PublicIndex) \times \dots \times \IsetDistribution_N(\Iset_\PublicIndex)$. A public belief state is a public state with associated range $\PublicBeliefState = (\Iset_\PublicIndex, \Range)$. Optimal value function is a function $\OptimalValueFunction(\PublicBeliefState)  \in \RealNumbers^{|\Isets_1(\Iset_\PublicIndex)| \dots |\Isets_{N}(\Iset_\PublicIndex)|}$, that for given public belief state $\PublicBeliefState$ returns counterfactual values $\CounterfactualValues^\Policy_\Player(\Iset_\Player)$ for each player $\Player$ and infoset $\Iset_\Player \in \Isets_\Player(\Iset_\PublicIndex)$ associated with public belief state $\PublicBeliefState$ if both players played optimally from that point based on associated ranges $\Range$. Value function $\ValueFunction(\PublicBeliefState)$ is then defined similarly, but it's output could be noisy. For any public belief state, we can construct subgame $\FOGGame(\PublicBeliefState)$, which is the same FOSG as $\FOGGame$, but it starts in $\Iset_\PublicIndex$ with initial ranges $\Range$ instead of the initial state. Depth-limited subgame is defined as $\FOGGame^{\text{DL}}(\PublicBeliefState)$. However, after some depth limit, the value function $\ValueFunction$ is called and returned counterfactual values are used as a terminal values. 

Resolving some subgame to improve blueprint policy $\Policy$ naively would involve fixing policy $\Policy$ for all players in previous parts of the game and using single $\ArtificalState$, where players act with corresponding ranges. However, this may result in a much more exploitable strategy \cite{cfrd} than the original blueprint $\Policy$. Multiple techniques were developed to deal with this issue in two-player zero-sum games \cite{cfrd,maxmargin,reachmaxmargin,rebel}. In this work we will focus on the one proposed in \cite{cfrd}. When resolving from the perspective of player $\Player$, it constructs a slightly modified subgame called a gadget game $\GadgetGame_\Player(\PublicBeliefState, \CounterfactualValues^\Policy_\Opponent)$, where $\CounterfactualValues^\Policy_\Opponent$ are counterfactual values for each infoset $\Iset_\Opponent \in \Isets_\Opponent(\Iset_\PublicIndex)$ of opponent $\Opponent$ in a current public state $\Iset_\PublicIndex$, when players play according to policy $\Policy$. This subgame has the artificial state $\ArtificalState$ as an initial state, but instead of using full ranges $\Range$, it uses only ranges $\Range_\Player$ and $\Range_\ChancePlayer$. Directly after this artificial state, $|\Histories(\Iset_\PublicIndex)|$ opponent's decision nodes are separated into multiple infosets based on the information player has in a given history. The opponent $\Opponent$ has a choice to either terminate the game and receive the counterfactual value $\CounterfactualValues^\Policy_{\Opponent}(\Iset_\Opponent)$ or continue the game as in the original subgame. 

Suppose that after the depth limit, the optimal value function is used and that the counterfactual values $\CounterfactualValues^{\Policy^{\NashEquilibrium}}_{\Opponent}(\Iset_\Opponent)$ used in the gadget game are computed for Nash equilibrium policy $\Policy^{\NashEquilibrium}$. Then the solution of gadget game $\GadgetGame_\Player(\PublicBeliefState, \CounterfactualValues^\Policy_\Opponent)$ is part of a Nash equilibrium in the entire game \cite{cfrd,valuefunctions}.

In this work, we focus on an alternative to the value function called multi-valued states \cite{MVS}, which, after the depth limit, gives the opponent the choice to fix its strategy until the end of the game. Then, the game terminates with the reward corresponding to the reward if player $\Player$ follows some blueprint policy $\Policy_\Player$, while the opponent follows the fixed strategy. Authors of \cite{valuefunctions} pointed out that multi-valued states could be viewed as a form of a value function. We further discuss this in \Cref{s:gameplay}

%\subsection{Reward regularization}
\subsection{Regularized Nash Dynamics} \label{s:rnad}
Using usual reinforcement learning techniques to find an optimal policy with self-play does not guarantee convergence to Nash equilibrium and can be exploited \cite{NFSP}. However, regularizing the reward leads to establishing this guarantee \cite{perolat2022mastering,sokotaregularization}. 

Reward regularization with KL-Divergence is 
\begin{align}
\begin{split}
    &\Rewards_\Player^\Regularized(\History, \Action, \Policy) := \\&\Rewards_\Player(\History, \Action) - \RegularizationTerm \log(\frac{\Policy(\Iset_\Player, \Action_\Player)}{\Policy^\Regularized(\Iset_\Player, \Action_\Player)})+ \RegularizationTerm \log(\frac{\Policy(\Iset_{-\Player}, \Action_{-\Player})}{\Policy^\Regularized(\Iset_{-\Player}, \Action_{-\Player})})
\end{split}
\end{align}
\begin{equation*}
\end{equation*}
where $\Policy^\Regularized$ is a regularization policy and $\RegularizationTerm$ is a hyperparameter for adjusting the regularization effect. Games with such a regularized reward have a single unique fixed point $\Policy_{\text{fix}}$ for any regularization policy $\Policy^\Regularized$ \cite{poincarre}. Follow the Regularized Leader \cite{followregularized} may be used to converge to $\Policy_{\text{fix}}$. Setting $\Policy^\Regularized = \Policy_{\text{fix}}$ will change the unique fixed point. If this is done repetitively, the fixed point will gradually converge to the Nash equilibrium.

In order to converge to a Nash equilibrium in large games with reward regularization, the authors in \cite{perolat2022mastering} propose using two-player V-trace with Neural Replicator Dynamics \cite{neurd} as a loss function for policy head and usual L2 loss for value head. With this method, it is possible to achieve human-level performance in Stratego. 

\section{\AlgorithmName}
In this section, we present an algorithm \AlgorithmName{} (\AlgorithmShort) that trains an additional critic alongside a policy network with some policy-gradient algorithm for imperfect information games, like RNaD \cite{perolat2022mastering}. During gameplay, this critic is used to construct a depth-limited gadget game that is then solved with some search algorithm, in our case CFR+ \cite{cfrplus}.
\subsection{Network Training} \label{s:network}
Performing a depth-limited search in imperfect information games requires a value function that, after the depth limit, assigns value to each infoset based on ranges \cite{valuefunctions}. To use a value function based on multi-valued states, it is necessary for each history $\History$ to learn multiple values that correspond to the expected values if the opponent plays some fixed policies against the player's blueprint policy. Training such value function presents two difficulties to overcome. The first is the selection of the opponent policies to be used as a basis to compute the expected values. In large games, simply using some predefined policy is not viable due to the sheer size of full policies. The second is estimating the expected value of these policies.

To find different opponent policies, we decided to use a similar method to the "bias approach" in \cite{MVS}. The main idea is to use a single policy, which in our case is the policy from the policy network, and then apply some mapping that maps the network policy to a different policy. We call this mapping a policy transformation.

\begin{definition} Policy transformation $\PolicyTransformation : \Policies_\Player \rightarrow \Policies_\Player$ for player $\Player$ is a function that maps policy to a different one. 
\end{definition}

In poker, an example of a transformation is folding more frequently \cite{Polaris,Abstractions}. Similarly, in Goofspiel, certain transformations increase the probability of playing cards with lower value if the player is winning or decrease the probability of playing even-valued cards. These transformations do not have to be game-domain specific. However, the effectiveness of different transformations varies across different games. These transformations aim to cover the strategy space as much as possible. Constant transformations that transform any policy into each pure strategy cover this space fully \cite{MVS}. 

In our experiments, we use transformations that move the actor policy $\pi$ in some direction with a fixed step size $k \in \mathbb{R}$. These transformations are encoded within a neural network, which undergoes concurrent training with the actor network. At each training step, for each trajectory, we compute the direction in which the actor policy for that trajectory has shifted. Subsequently, we update only the transformation associated with the minimum Euclidean distance across the trajectory. Since each update should improve the actor policy against the current opponent's policy, this direction should aim into the important regions of strategy space, but the quality of these transformations is not guaranteed.

To estimate the expected value of these transformed policies against the player's blueprint, we propose to train an additional critic network $\CriticNetwork_\Player(\History)[\PolicyTransformation]$, that takes history $\History$ as an input and outputs the player's $\Player$ expected value for each policy transformation $\PolicyTransformation$. To avoid sampling additional trajectories, which may be impossible in continuing tasks, we would like to use already sampled trajectories by the original policy-gradient algorithm more effectively. This may be achieved by using some off-policy training algorithm. In this work, we have used a V-trace estimator.

We define the V-trace operator $\VtraceOperator$ for histories, while following some sampling policy $\SamplingPolicy$ as
\begin{align}
\begin{split}
    &\VtraceOperator \VtraceVal_\Player(\History) = \VtraceVal_\Player(\History) +\\ & \ExpectedValue_\SamplingPolicy \left[ \sum^\TrajectoryLength_{\timestep = 0}  \rho^\timestep \Big( \prod_{\mathclap{0 \leq \smallertimestep < \timestep}} \DiscountFactor c^\smallertimestep \Big) \Big(\Reward_\Player^\timestep + \DiscountFactor \VtraceVal_\Player(\History^{\timestep + 1}) - \VtraceVal_\Player(\History^\timestep)\Big)| \History^0 = \History, \SamplingPolicy \right]
    \end{split}
\end{align}
where $\History^\timestep$ is history at timestep $\timestep$ in trajectory $\tau$ with initial history being $\History^{0}$. $\rho^\timestep = \min (\overline \rho, \frac{\Policy(\Iset_\Player(\History^\timestep), \Action_\Player^\timestep)}{\SamplingPolicy(\Iset_\Player(\History^\timestep), \Action_\Player^\timestep)})$ and $c^\timestep = \min (\overline c, \frac{\Policy(\Iset_\Player(\History^\timestep), \Action_\Player^\timestep)}{\SamplingPolicy(\Iset_\Player(\History^\timestep), \Action_\Player^\timestep)})$ are truncated importance sampling weights. If player does not act in timestep $\timestep$, or if the $\timestep \geq l$, we say that $\VtraceVal_\Player(\History^\timestep) = 0$.

\begin{theorem} \label{thm:vtrace} (Inspired by \cite{Vtrace}): Let $\rho^\timestep = \min{ (\overline \rho, \frac{\Policy(\Iset_\Player(\History^\timestep), \Action_\Player^\timestep)}{\SamplingPolicy(\Iset_\Player(\History^\timestep), \Action_\Player^\timestep)} )}$, $c^\timestep = \min{ (\overline c, \frac{\Policy(\Iset_\Player(\History^\timestep), \Action_\Player^\timestep)}{\SamplingPolicy(\Iset_\Player(\History^\timestep), \Action_\Player^\timestep)} )}$, $\overline \rho \geq \overline c \geq 1$ and $\SamplingPolicy(\Iset, \Action) > 0 \;\; \forall \Iset, \Action$. Let us assume there exists $\ContractionProofConstant \in (0, 1]$, such that $\ExpectedValue_\SamplingPolicy \rho^{\TrajectoryLength} \Big( \prod_{{0 \leq \smallertimestep < \TrajectoryLength}} c^\smallertimestep \Big) \geq \ContractionProofConstant$, where $l$ is a length of the trajectory. Then the V-trace operator $\VtraceOperator$ has a unique fixed point $\VtraceVal^{\Policy_{\overline \rho}}$, which is the expected value of following policy
\begin{align}
\begin{split}
\Policy_{\overline \rho}(&\Iset_\Player(\History^\timestep), \Action_\Player^\timestep) =\\& \frac{\min (\overline \rho \SamplingPolicy(\Iset_\Player(\History^\timestep), \Action_\Player^\timestep), \Policy (\Iset_\Player(\History^\timestep), \Action_\Player^\timestep))}{\sum_{b \in \Actions_\Player(\Iset_\Player(\History^\timestep))} \min (\overline \rho \SamplingPolicy(\Iset_\Player(\History^\timestep), b), \Policy (\Iset_\Player(\History^\timestep), b))}
\end{split}
\end{align}
\end{theorem}

Using $\overline \rho \rightarrow \infty$ will result in convergence to the $\VtraceVal^\Policy$, that is, the expected value of the policy $\Policy$ in given history. The proof of \Cref{thm:vtrace} is in Appendix \ref{app:proofs}. For training the history network, we use standard L2 loss. %We provide pseudocode for training in Appendix \ref{app:pseudo}.

\subsection{Constructing Public Belief State} \label{s:gameplay}
After the training phase, the original policy-gradient algorithm ends up with a trained policy network, which is then used at each decision point to retrieve policy. Based on this policy, the action is sampled, and the game proceeds. \AlgorithmShort{} finishes the training with the additional history value network.

As discussed in the \Cref{s:motivation}, naively employing search in the game could lead to highly exploitable strategies. Using resolving gadget \cite{cfrd}, which limits the opponent's counterfactual value in root infosets, bounds the error after performing search in a subgame. Constructing a gadget game requires the opponent's counterfactual values $\CounterfactualValues^\Policy_\Opponent$ in each root infoset and public belief state $\PublicBeliefState = (\Iset_\PublicIndex, \Range)$. Most previous methods stored counterfactual values $\CounterfactualValues^\Policy_\Opponent$, public state $\Iset_\PublicIndex$ and ranges $\Range$ from previous searches \cite{deepstack,pog}. \AlgorithmShort{} could leverage the same idea. Still, we propose a novel approach of using the trained history network to compute counterfactual values, the trained policy network to retrieve ranges for searching player $\Player$, and a function that reconstructs the public state. This allows running the search from an arbitrary public state without the need to search in all parent public states.

Safe search in imperfect information games has to be performed from all possible histories based on public information. This requires all histories $\Histories(\Iset_\PublicIndex)$ from current public state $\Iset_\PublicIndex$ to construct the gadget game $\GadgetGame_\Player$. Let us assume a function that generates all histories in the public state $\Iset_\PublicIndex$ from the publicly available information to all players. While the details of reconstructing the public state $\Iset_\PublicIndex$ are not explored in detail here, for games like Goofspiel or Battleships, the public state can be reconstructed by Constraint Satisfaction Problem (CSP) \cite{modernapproach,CSPGS}.

Gadget game $\GadgetGame_\Player$ requires ranges for the player $\Player$ and the chance player. Chance player ranges $\Range_\ChancePlayer$ can easily be stored during gameplay like older methods \cite{deepstack,pog,libratus}. Suppose the player $\Player$ played only by using the policy from the neural network during current gameplay and decides to perform a search at some public state $\Iset_\PublicIndex$. In that case, it has to recompute the ranges $\Range_\Player$ for each infoset $\Iset_\Player \in \Isets_\Player(\Iset_\PublicIndex)$. This requires backtracking from all of these infosets to all previous decision points and acquiring the policy from the policy network for each of them. Because we are in a perfect recall setting, each infoset cannot have more parenting infosets than the maximal trajectory length $\TrajectoryLength^{\text{max}}$ to this infoset from the initial state $\InitState$, so this backtracking involves at most $|\Isets_\Player(\Iset_\PublicIndex)| \TrajectoryLength^{\text{max}}$ queries to the policy network.
%During the gameplay, at each decision point, the agent has a choice to either play according to the network policy as with the original algorithm or use more computational resources to refine the strategy with additional search. This choice is completely independent of using search in any previous parts of the game, unlike some previous methods, which used continual resolving \cite{cfrd, deepstack, pog, libratus, MVS}. 

%These methods used stored counterfactual values $\CounterfactualValues^\Policy_\Opponent$  and public belief state $\PublicBeliefState$ from previous resolves during gameplay to construct a gadget game. Since \AlgorithmShort{} does not require these resolves in previous parts of the game, we require to have a function that would reconstruct the public state. 

In order to compute the counterfactual values in root infosets for the opponent, let us assume that one of the transformations we used in training is identity transformation, meaning that the critic network output corresponds to the expected value if both players follow the policy from the network, which serves as an approximation of a Nash equilibrium. The opponent's counterfactual values are the expected values multiplied by the reaches of both chance and resolving players. As discussed, these ranges are always available, so the counterfactual values can always be computed from the definition. Thanks to this, we have both counterfactual values $\CounterfactualValues^\Policy_\Opponent$ and $\PublicBeliefState$ available in each decision point to construct a gadget game and perform a safe search.

We compare previously used continual resolving for constructing public belief state with our approach in \Cref{s:addcfv}
\subsection{Value Function} \label{s:valuefunctions}
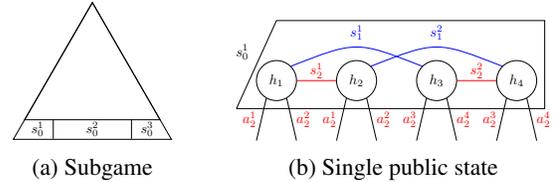
\begin{figure}[h]
    \centering
  \begin{subfigure}{.14\textwidth}
      \centering
        \resizebox{2.4cm}{!}{
    \begin{tikzpicture}
        \node[isosceles triangle,
        isosceles triangle apex angle=60,
            draw,
            rotate = 90,
            anchor=apex,
            minimum size =2.5cm] (Trunk)at (0,0){};
        \node[isosceles triangle,
        isosceles triangle apex angle=60,
            draw,
            rotate = 90,
            anchor=apex,
            minimum size =3.5cm] (TrunkMVS)at (0,0){};
        
        \draw (-0.8,-2.5) -- (-0.8, -3.5);
        \draw (0.8,-2.5) -- (0.8, -3.5);
        \draw (-1.3,-3) node {\huge $s_0^1$};
        \draw (0,-3) node {\huge $s_0^2$};
        \draw (1.25,-3) node {\huge $s_0^3$};
    \end{tikzpicture} 
    }
      \caption{Subgame}
      \label{fig:vfsubgame}
  \end{subfigure}
  \begin{subfigure}{.3\textwidth}
      \centering
        \resizebox{5.1cm}{!}{
\begin{tikzpicture}

% \node[rectangle, draw,  minimum height=2.1cm, minimum width=7.5cm] (S0) at (3,0.4) {};
\draw[] (-1, -0.7) -- (6.7, -0.7) -- (6.7, 1.8) -- (0, 1.8) -- node[left=1em, above] {\LARGE $s_0^1$} (-1, -0.7);

% \node[anchor=north west,inner sep=3pt] at (S0.north west) 
%     {$s_0^1$};
\node[circle,draw, minimum size=1cm] (W0) at  (0,0) {\LARGE $h_1$};
\node[circle,draw, minimum size=1cm] (W1) at  (2,0)  {\LARGE $h_2$};
\node[circle,draw, minimum size=1cm] (W2) at  (4,0)  {\LARGE $h_3$};
\node[circle,draw, minimum size=1cm] (W3) at  (6,0)  {\LARGE $h_4$};
\draw[red] (W0.east) -- (W1.west); 
\draw[red] (W2.east) -- (W3.west); 
\draw[blue] (W0.north east) .. controls (2, 1.2).. (W2.north west); 
\draw[blue] (W1.north east) .. controls (4, 1.2).. (W3.north west); 
\draw[blue] (2, 1.4) node {\LARGE$s_1^1$};
\draw[blue] (4, 1.4) node {\LARGE$s_1^2$};
\draw[red] (1, 0.35) node {\LARGE$s_2^1$};
\draw[red] (5, 0.35) node {\LARGE$s_2^2$};
\node[] (W0leftbottom)  at (W0.240) {};
\node[] (W0rightbottom)  at (W0.300) {};
\node[] (W1leftbottom)  at (W1.240) {};
\node[] (W1rightbottom)  at (W1.300) {};
\node[] (W2leftbottom)  at (W2.240) {};
\node[] (W2rightbottom)  at (W2.300) {};
\node[] (W3leftbottom)  at (W3.240) {};
\node[] (W3rightbottom)  at (W3.300) {};
% \node[] (W0leftbottom)  at (W0.240) {};
% \node[draw]  (W0bottom) at ()(W.360/17) {};
\draw (W0leftbottom.center) -- node[left, red]{\LARGE$a_2^1$} (-0.5, -1.7) ;
\draw (W0rightbottom.center) -- node[right, red]{\LARGE$a_2^2$} (0.5, -1.7) ;
\draw (W1leftbottom.center) -- node[left, red]{\LARGE$a_2^1$} (1.5, -1.7) ;
\draw (W1rightbottom.center) -- node[right, red]{\LARGE$a_2^2$} (2.5, -1.7) ;
\draw (W2leftbottom.center) -- node[left, red]{\LARGE$a_2^3$} (3.5, -1.7) ;
\draw (W2rightbottom.center) -- node[right, red] {\LARGE$a_2^4$} (4.5, -1.7) ;
\draw (W3leftbottom.center) -- node[left, red]{\LARGE$a_2^3$} (5.5, -1.7) ;
\draw (W3rightbottom.center) -- node[right, red]{\LARGE$a_2^4$} (6.5, -1.7);
\end{tikzpicture}
    }
      \caption{Single public state}
      \label{fig:vfps}
  \end{subfigure}
  \caption{Subgame with 3 leaf public states. Right part shows the detail of public state $\Iset_\PublicIndex^1$ with 4 histories and 2 information sets for each player. The multi-valued states technique modifies this public state by giving player 2 a choice between two strategies in each history against blueprint policy of player 1.}
    \label{fig:vf}
\end{figure}
Multi-valued states \cite{MVS} are a modification of the original game, where after the depth limit, the opponent has one more decision point, in which it chooses its strategy for the remainder of the game against some fixed blueprint policy. This modified game is a standard game and can be solved by any game solving algorithm unaware of value functions. Authors of \cite{deepstack} trained value function for Poker, which served as a generalization of the value function from perfect information games. This value function returns counterfactual values based on ranges at the depth limit. These were then specially adapted for Counterfactual Regret Minimization (CFR) \cite{OriginalCFR}.

These two approaches to the depth-limited search may not seem connected, but the authors of \cite{valuefunctions} show this connection by explaining that the multi-valued states could be viewed as a value function, which is not trained for all possible strategies in the game, but only for some subset. However, their explanation does not explicitly clarify how this connection could be leveraged. 

Let us assume a depth-limited subgame, as in \Cref{fig:vf}, with three leaf public states $\Iset_\PublicIndex^1, \Iset_\PublicIndex^2, \Iset_\PublicIndex^3$. Further examining public state $\Iset_\PublicIndex^1$ in \Cref{fig:vfps}, there are four possible states in this public state, which are separated into different infosets. First player has infoset $\Iset_1^1$, that contains histories $\History_1, \History_3$ and $\Iset_1^2$ containing histories $\History_2, \History_4$. Similarly second player has infosets $\Iset_2^1, \Iset_2^2$, that contains histories $\History_1, \History_2$ and $\History_3, \History_4$ respectively. The value function would return counterfactual values for each infoset, namely $\CounterfactualValues^\Policy_1(\Iset_1^1), \CounterfactualValues^\Policy_1(\Iset_1^2), \CounterfactualValues^\Policy_2(\Iset_2^1), \CounterfactualValues^\Policy_2(\Iset_2^2)$. 

In the same example, using multi-valued states as proposed in \cite{MVS} gives the opponent, player 2, another choice to pick between two strategies in each infoset, as shown in \Cref{fig:vfps}.  

Using CFR requires storing multiple statistics in each infoset, like regret and average strategy. These statistics must be stored even for these artificial infosets created by the multi-valued states. In such a scenario, the multi-valued states could be viewed as a stateful value function where these statistics determine the value function state. Each CFR iteration, this value function updates its state and returns counterfactual values based on the current state and input ranges $\Range$. Let us assume that instead of performing regret matching and storing these necessary statistics, the opponent always picks a best response based on input ranges. In such a case, we can use multi-valued states as a usual value function with just additional computation step, without keeping the state inside the value function. Let us rephrase the value function in terms of using a history network with policy transformations for player $\Player$.
\begin{equation}
    \ValueFunction(\PublicBeliefState)[\Iset_\Player] = \min_{\PolicyTransformation \in \PolicyTransformations} \frac{\sum_{\History \in \Histories(\Iset_\Player)}\PlayerReach^{\Policy}_{-\Player}(\History)  \CriticNetwork_\Player(\History)[ \PolicyTransformation]}{\sum_{\History \in \Histories(\Iset_\Player)} \PlayerReach^{\Policy}_{-\Player}(\History)} 
\end{equation}
where $\PolicyTransformations$ is a set of all used policy transformations and $\PlayerReach^{\Policy}_{-\Player}$ are directly taken from ranges $\Range_{-\Player}$ in $\PublicBeliefState$. Since we are in a two-player zero-sum setting and the network returns an expected value for the player $\Player$, choosing a best response by the opponent is taking such transformation that minimizes the counterfactual value of the player $\Player$.

This presents a new way to train value functions without using search in the training time at all by first training multi-valued states and then using them with different possible ranges to generate counterfactual values for training the value function.

With the constructed gadget game and a corresponding value function, any previously developed search algorithm for imperfect information games may conduct the subsequent search. During our experiments, we used CFR+ \cite{cfrplus} exclusively. When using CFR+, a subclass of Growing tree-CFR (GT-CFR) introduced in \cite{pog}, the bounds already established by the GT-CFR authors in Theorem 3 also apply to our approach. However, the quality of this value function $\xi$, which is the distance between values returned by the optimal value function and the trained value function, is affected by additional errors, even if the value function is trained perfectly. The first of these errors is that the blueprint strategy of resolving player $\Player$ is not Nash equilibrium but just some approximation. Second is that the opponent's strategies used for multi-valued states may not be best responses to any strategy.

The multi-valued states could be expanded by giving both players a choice to pick from multiple strategies for the rest of the game \cite{pluribus,valuefunctions}. This would increase the number of values that must be trained quadratically because each pair of strategies between players yields different values. \AlgorithmShort{} is compatible with this approach, and the only required change is to compute a Nash equilibrium of the underlying normal-form game at the depth limit during search instead of a best response.

We have already shown how the history critic trained for multi-valued states can be used as a value function. However, we could go one step further and train the usual value function out of this history critic. The value function may be trained before the gameplay phase by randomly sampling possible ranges, computing the counterfactual values from multi-valued states, and then training the value function on these counterfactual values. Such trained value function can then be used in already developed algorithms \cite{deepstack,pog} without any changes. 

We provide pseudocode for training and gameplay in \Cref{app:pseudocodes}
% We have already shown how the history critic trained for multi-valued states can be used as a value function. However, we could go one step further and train the usual value function out of this history critic. This would completely avoid the need to call the critic for each history and further compute a best response or Nash equilibrium when using multi-valued states for single or both players. The value function may be trained before the gameplay phase by randomly sampling possible ranges, computing the counterfactual values, and then training the value function on these counterfactual values. Such trained value function can then be used in already developed algorithms \cite{deepstack,pog} without any changes. The quality of such a value function depends on the chosen transformations when training multi-valued states. In our work, we have always used the history critic and have done computation of a best response during the search.

%The multi-valued states could further be expanded by not assuming that the resolving player has just one strategy to play, but it could also pick between multiple of them. Since both players can choose a strategy to play, the resulting value would not be computed as a best response, but as a Nash Equilibrium of a corresponding normal-form game. 

%When viewing value function from this point of view, it is obvious that the output of imperfect information value function should be counterfactual values, if both players play optimally from given public belief state $\PublicBeliefState$. It also shows

% \subsection{Advantages of \AlgorithmName}

\section{Experiments} \label{s:experiments}
We have conducted two experiments to validate \AlgorithmShort{}'s performance. Our experiments have used the framework OpenSpiel \cite{OpenSpiel}. The depth-limited search was conducted with CPU Intel Xeon Scalable Gold 6146, operating at a frequency of 3.2 GHz, while the network training additionally used a single GPU Tesla V100. The hyperparameter setting is described in Appendix \ref{app:details}. Appendix \ref{app:gamerules} provides a detailed description of the game rules.
\subsection{Head-To-Head Play in Large Games} \label{s:largeexps}
The first experiment shows the effectiveness of \AlgorithmShort{} in large games. We trained RNaD and RNaD with \AlgorithmShort{} separately, each three times, with the same hyperparameter setting. These were then used in head-to-head play against each other, where RNaD directly followed the policy network and \AlgorithmShort{} applied search in those subgames that contained less than 25000 or 100000 unique histories. The evaluation encompassed these games: Goofspiel with eight cards and randomized point order, and with thirteen cards both with descending and randomized point order and Battleships on square grids of size 3x3, 5x5, and 7x7 with ships of size (2, 2), (3, 2, 2), (4, 3, 3, 2) respectively. % Goofspiel with eight cards and descending order of point cards, which contains approximately $2 \cdot 10^9$ unique histories; Goofspiel with eight cards and randomized order of point cards, having $9 \cdot 10^{13}$ unique histories; Goofspiel with thirteen cards and descending order of point cards, containing approximately $5 \cdot 10^{19}$ unique histories; Goofspiel with thirteen cards and randomized order of point cards, having roughly $3 \cdot 10^{29}$ unique histories; Battleships on a square grid of size 3 with 2 ships of sizes 2 and 2 containing roughly $10^{14}$ unique histories and Battleships on square grid of size 5 with 3 ships of sizes 3,2 and 2 that have roughly $10^{57}$ unique histories.

For each game, we have trained 3 RNaD policy networks and 3 different RNaD policy networks along with \AlgorithmShort{}'s history critic. Each network was trained for 3 days. We have run over one hundred thousand game simulations of RNaD against \AlgorithmShort{} for each game. In half of them, the \AlgorithmShort{} was a first player; in half, it was a second player. The average rewards multiplied by 100 from these simulations are presented in \Cref{tab:heads} along with a 95\% confidence interval. Employing search outperformed the RNaDs policy network with statistical significance in all the games tested. 

The usage of search improves the performance in Battleships more than in Goofspiel. This is because, in Goofspiel, most of the subgames are quite large, so the search is used sparingly. However, Battleships have a lot of smaller subgames where the search is used.% Moreover, the Battleships is a much larger game than the Goofspiel, and similar subgames in Battleships may have vastly different optimal policies. Therefore, training RNaD in Battleships is harder than in Goofspiel.

\begin{table}[h]
\centering
        \resizebox{8.5cm}{!}{
\begin{tabular}{c|c|c|c|c|c|c}
\begin{tabular}[c]{@{}c@{}}Max subgame\\ size\end{tabular} & \begin{tabular}[c]{@{}c@{}}Goofspiel 8\\ randomized\end{tabular} & \begin{tabular}[c]{@{}c@{}}Goofspiel 13\\ descending\end{tabular} & \begin{tabular}[c]{@{}c@{}}Goofspiel 13\\ randomized\end{tabular}& \begin{tabular}[c]{@{}c@{}}Battleships\\3x3\end{tabular} & \begin{tabular}[c]{@{}c@{}}Battleships\\5x5\end{tabular} & \begin{tabular}[c]{@{}c@{}}Battleships\\7x7\end{tabular} \\ \hline
25000                                                                                                 & $2.70 \pm 0.05$                                                   & $3.44 \pm 0.05$                                                    & $1.12\pm 0.05$                                                    & $6.2 \pm 0.06$                                               & $81.9 \pm 0.3$                & $24.1 \pm 0.7$                                   \\
100000                                                                                       & $3.10 \pm 0.19$                                                   & $3.51 \pm 0.1$                                                    & $1.43 \pm 0.3$                                                     & $6.5 \pm 0.07$                                              & $83.5 \pm 0.5$    & $42.1 \pm 0.9$   
\end{tabular}
}
\caption{Average rewards from head-to-head play of RNaD against RNaD with \AlgorithmShort. The average rewards are multiplied by 100.}
\label{tab:heads}
\end{table}
\subsection{Exploitability in Smaller Games} \label{s:smallexps}
\begin{figure}[h!]
\centering
  \begin{subfigure}{.237\textwidth}
      \centering
      \includegraphics[width=.96\linewidth]{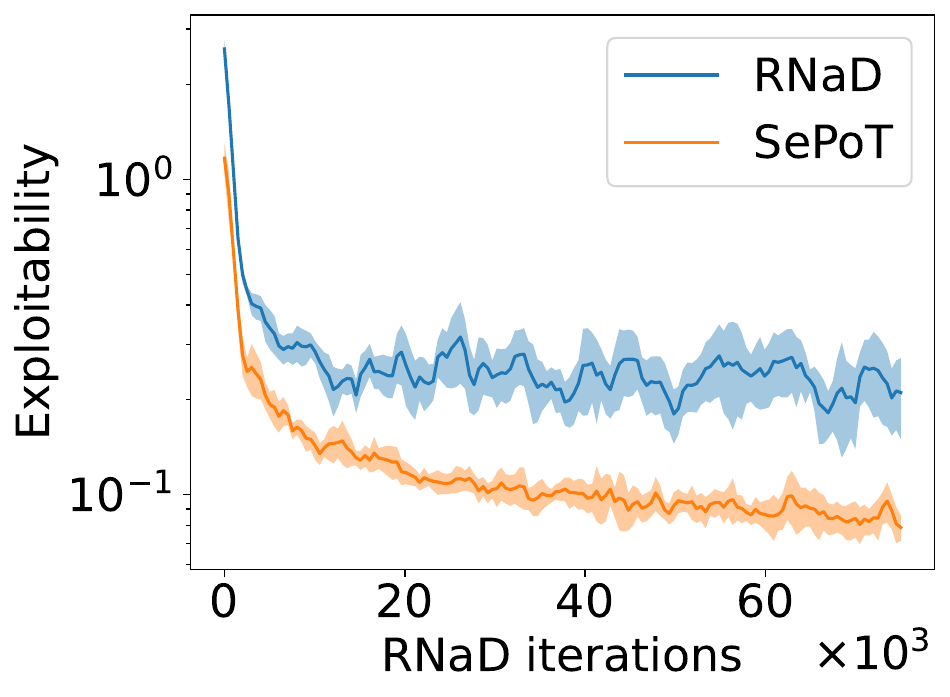}
      \caption{Leduc hold'em}
      \label{fig:leducexploit}
  \end{subfigure}
  \begin{subfigure}{.237\textwidth}
      \centering
      \includegraphics[width=.96\linewidth]{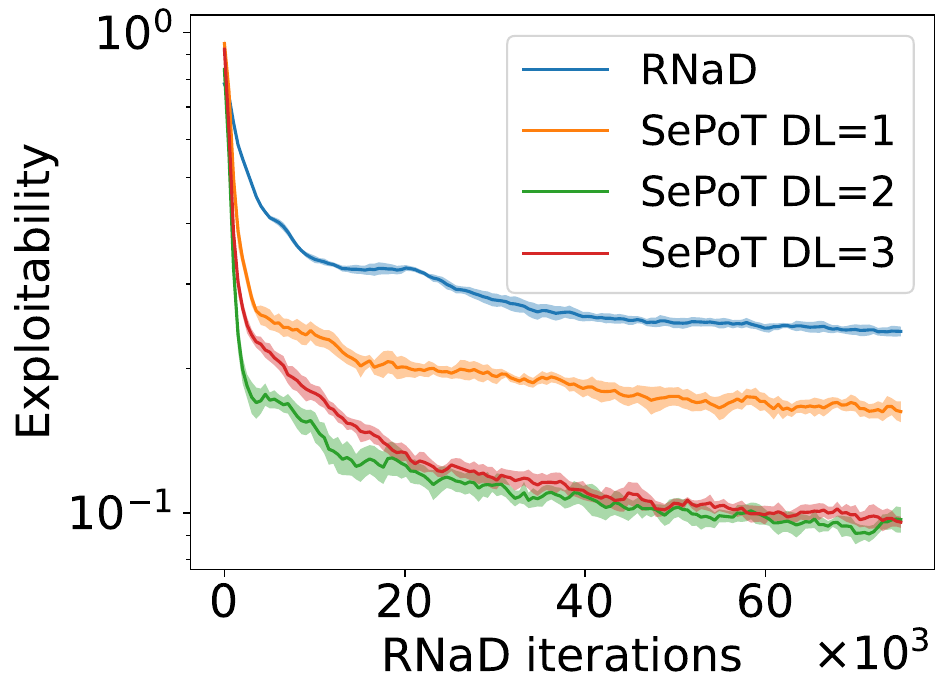}
      \caption{Goofspiel}
      \label{fig:goofspielexploit}
  \end{subfigure}
  \caption{Exploitability of policy network RNaD and the search with \AlgorithmShort{} based on RNaD training iterations.}
  \label{fig:exploitexperiment}
  % \Description{Exploitability}
\end{figure}

In the second experiment, we aim to evaluate the effect of depth-limited search with multi-valued states on the quality of the policy compared to the policy network. In the experiments, we compare the exploitability of \AlgorithmShort{} and the policy network from RNaD. We use Leduc hold'em and Goofspiel with 5 cards and randomized order of point cards for evaluation. We trained ten networks with different random seeds while computing exploitability every 500 iterations up to iteration 75,000. The results in \Cref{fig:exploitexperiment} show the mean and 95\% confidence interval of the exploitability throughout these runs. In Leduc hold'em, we separately solve part before dealing a public card. In Goofspiel, we use different depth limits for search, where DL=$n$ means that both players play $n$ actions before the value function is called.

The plots in \Cref{fig:goofspielexploit} show how the exploitability changes with additional iterations. Using the depth limit 2 or 3 yields similar results, showing that both outperform depth limit 1 and the policy network. This experiment shows that with additional training, the search still improves over the network policy. Furthermore, it confirms the intuition that having a longer look-ahead increases the quality of the search solution.
\section{Related Work} \label{s:related}
Previous search algorithms require using safe search both in train and test time \cite{rebel,libratus,deepstack,pog}. Using those algorithms in larger games requires some subgame reduction techniques like abstracting the game \cite{libratus} or revealing part of the private information \cite{kls}. \AlgorithmShort{} uses a policy-gradient training algorithm that trains in arbitrarily sized games. Furthermore, \AlgorithmShort{} chooses whether the search should be used at each point independently during gameplay. This enables safe search in larger games than previously possible without using any subgame reduction techniques. However, \AlgorithmShort{} is flexible enough to be used with these subgame reduction techniques to increase the number of subgames that the search may be applied in even further. 

Imagine a game of Battleships as an example. After placing the ships, there are more than $10^{20}$ possible states in a public state, which makes training of search algorithms infeasible. After several more rounds, the information is steadily being revealed to both players, and eventually, the size of the public state becomes small enough. \AlgorithmShort{} can perform search in such a game, which was previously not possible. 

Our method is heavily based on the \cite{MVS}, with the notable differences being a more principled way to train the blueprint policy with policy-gradient algorithms and choosing the policy transformations. Moreover, we show how to use multi-valued states as a value function practically. We present a way to leverage an already trained critic network to construct a gadget game anywhere in the game without the need to perform a search in previous parts.

Model-free deep reinforcement learning methods based on regret minimization \cite{DREAM,ARMAC,ESCHER} use search in the gameplay and still can be scaled to very large games. The main limitation of these methods is the need to compute the average strategy, which either requires storing all the trajectories used in learning or approximating this average, which is prone to error. In games like Battleships, the training may consist of billions of trajectories to train properly. Storing all of these trajectories and computing averages is both memory and computationally demanding. \AlgorithmShort{} uses a policy-gradient algorithm that approximates the optimal strategy in immediate strategies, so it does not compute the average strategy.
% Effective reconstruction of the public belief state was proposed by the authors in \cite{beliefstatemodeling}
% Algorithms for imperfect information games outside of two-player zero-sum scope like Pluribus for multiplayer Poker \cite{pluribus}, Sparta for cooperative game Hanabi \cite{hanabi} or Cicero for the multiplayer adversarial game with cooperative elements Diplomacy \cite{diplomacy} all used search to achieve strong performances. These advancements underscore the significance of search as an integral component in gameplaying algorithms.
\section{Conclusion} \label{s:conclusion}
Policy-gradient algorithms proved to be powerful for solving huge problems in various settings. Recently, these algorithms were successfully applied to two-player zero-sum imperfect information games and outperformed human professionals in Stratego. However, these algorithms could not incorporate search to improve mistakes caused by network inaccuracies. We have presented an algorithm \AlgorithmShort{} that can be used alongside any trajectory sampling policy-gradient algorithm in two-player zero-sum imperfect information games to train additional critic network without any search during training. We have shown that this network is sufficient to enable safe search during test time. The search we use differs from previous approaches since it is independent of the application of search in previous parts of the game, with the only limitation being the local size of the subgame and reconstruction of the public states. Thanks to this, \AlgorithmShort{} can employ search even in larger games while still having a choice to follow policy from the policy network. Our experimental results demonstrate that \AlgorithmShort{} is able to outperform RNaD in most of the tested scenarios.

%% file: appendix.tex
\appendix
\section{Proofs} \label{app:proofs}
This proof of \Cref{thm:vtrace} is based on the proof of the original V-trace \cite{Vtrace}, with the notable difference that we discuss only the case with finite horizon and discount factor $\gamma = 1$.
We can reformulate the V-trace operator as
\begin{align*}
    \VtraceOperator \VtraceVal_\Player&(\History) = \VtraceVal_\Player(\History)\\ &+ \ExpectedValue_\SamplingPolicy \left[ \sum^\TrajectoryLength_{\timestep = 0}  \rho^\timestep \Big( \prod_{0 \leq \smallertimestep < \timestep} \DiscountFactor c^\smallertimestep \Big) \Big(\Reward_\Player^\timestep + \DiscountFactor \VtraceVal_\Player(\History^{\timestep + 1}) - \VtraceVal_\Player(\History^\timestep) \Big) \right] \\
     &= (1 - \ExpectedValue_\SamplingPolicy \rho^0) \VtraceVal_\Player(\History)\\
     &+ \ExpectedValue_\SamplingPolicy \left[ \sum^\TrajectoryLength_{\timestep = 0} \Big( \prod_{0 \leq \smallertimestep < \timestep} \DiscountFactor c^\smallertimestep \Big) \Big( \rho^\timestep \Reward_\Player^\timestep + [ \rho^\timestep - c^\timestep \rho^{\timestep + 1}] \DiscountFactor \VtraceVal_\Player(\History^{\timestep + 1}) \Big) \right] 
\end{align*}
Now we can find the contraction coefficients
\begin{align*}
    &\VtraceOperator \VtraceVal_\Player(\History)  - \VtraceOperator \overline \VtraceVal_\Player(\History) =\\
    &= (1 - \ExpectedValue_\SamplingPolicy \rho^0) (\VtraceVal_\Player(\History) - \overline \VtraceVal_\Player(\History))\\
    &+ \ExpectedValue_\SamplingPolicy \left[ \sum^\TrajectoryLength_{\timestep = 0} \DiscountFactor \Big( \prod_{0 \leq \smallertimestep < \timestep} \DiscountFactor c^\smallertimestep \Big) \Big( [ \rho^\timestep - c^\timestep \rho^{\timestep + 1}] [ \VtraceVal_\Player(\History^{\timestep + 1}) - \overline \VtraceVal_\Player(\History^{\timestep + 1})] \Big) \right] \\
    &= \ExpectedValue_\SamplingPolicy \left[ \sum^\TrajectoryLength_{\timestep = 0} \DiscountFactor \Big( \prod_{0 \leq \smallertimestep < \timestep - 1} \DiscountFactor c^\smallertimestep \Big) \Big( [ \rho^{\timestep -1} - c^{\timestep  - 1 } \rho^{\timestep}] [ \VtraceVal_\Player(\History^{\timestep}) - \overline \VtraceVal_\Player(\History^{\timestep})] \Big) \right] 
\end{align*}
where $\rho^{-1} = c^{-1} = 1$ and ${0 \leq \smallertimestep < \timestep - 1} \DiscountFactor c^\smallertimestep = 1$ for $t \in {0, 1}$. Now we can show that all coefficient are non-negative, if $\overline \rho \geq \overline c$
\begin{equation*}
    \ExpectedValue_\SamplingPolicy \Big[  \rho^{\timestep - 1} - c^{\timestep - 1} \rho^{\timestep}\Big] \geq \ExpectedValue_\SamplingPolicy \Big[ c^{\timestep - 1} ( 1 - \rho^\timestep) \Big] \geq 0
\end{equation*}
In order to show that the V-trace operator is a contraction, we have to show that coefficients are strictly less than 1.
\begin{align*}
&\ExpectedValue_\SamplingPolicy \left[ \sum^\TrajectoryLength_{\timestep = 0} \DiscountFactor \Big( \prod_{0 \leq \smallertimestep < \timestep - 1} \DiscountFactor c^\smallertimestep \Big) \Big( [ \rho^{\timestep -1} - c^{\timestep  - 1 } \rho^{\timestep}] \Big) \right]= \\
=&\sum^\TrajectoryLength_{\timestep = 0} \DiscountFactor  \ExpectedValue_\SamplingPolicy \left[  \Big( \prod_{0 \leq \smallertimestep < \timestep - 1} \DiscountFactor c^\smallertimestep \Big)  \rho^{\timestep -1} \right] - \sum^\TrajectoryLength_{\timestep = 0} \DiscountFactor \ExpectedValue_\SamplingPolicy \left[  \Big( \prod_{0 \leq \smallertimestep < \timestep} \DiscountFactor c^\smallertimestep \Big) \rho^{\timestep} \right]\\
 =&\sum^\TrajectoryLength_{\timestep = 0} \DiscountFactor  \ExpectedValue_\SamplingPolicy \left[  \Big( \prod_{0 \leq \smallertimestep < \timestep - 1} \DiscountFactor c^\smallertimestep \Big)  \rho^{\timestep -1} \right] \\ &- \sum^\TrajectoryLength_{\timestep = 0} \DiscountFactor \ExpectedValue_\SamplingPolicy \left[  \Big( \prod_{0 \leq \smallertimestep < \timestep} \DiscountFactor c^\smallertimestep \Big) \rho^{\timestep} \right] + \rho^{-1} - \rho^{-1} \\
=& \sum^\TrajectoryLength_{\timestep = 0} \DiscountFactor  \ExpectedValue_\SamplingPolicy \left[  \Big( \prod_{0 \leq \smallertimestep < \timestep - 1} \DiscountFactor c^\smallertimestep \Big)  \rho^{\timestep -1} \right]\\ &-  \sum^{\TrajectoryLength + 1}_{\timestep = 0} \DiscountFactor \ExpectedValue_\SamplingPolicy \left[  \Big( \prod_{0 \leq \smallertimestep < \timestep - 1} \DiscountFactor c^\smallertimestep \Big) \rho^{\timestep - 1} \right] + 1 \\
=& 1  - \ExpectedValue_\SamplingPolicy \left[  \Big( \prod_{0 \leq \smallertimestep < \TrajectoryLength} \DiscountFactor c^\smallertimestep \Big)  \rho^{\TrajectoryLength} \right] \leq  1 - \ContractionProofConstant < 1
\end{align*}
Therefore $||\VtraceOperator  \VtraceVal_\Player(\History) - \VtraceOperator  \overline \VtraceVal_\Player(\History)|| \leq \ContractionConstant ||\VtraceVal_\Player - \overline \VtraceVal_\Player||_\infty$, where $\ContractionConstant = 1  - \ExpectedValue_\SamplingPolicy \left[  \Big( \prod_{0 \leq \smallertimestep < \TrajectoryLength} \DiscountFactor c^\smallertimestep \Big)  \rho^{\TrajectoryLength} \right] \leq  1 - \ContractionProofConstant < 1$
Let us now show what is the fixed point of $\VtraceOperator$
\begin{align*}
    &\ExpectedValue_\SamplingPolicy \rho^\timestep \Big(\Reward_\Player^\timestep + \DiscountFactor \VtraceVal^{\Policy_{\overline \rho}}_\Player(\History^{\timestep + 1}) - \VtraceVal^{\Policy_{\overline \rho}}_\Player(\History^\timestep) |\History^\timestep \Big)\\
    =& \sum_{\Action_\Player \in \Actions_\Player(\History^\timestep)} \SamplingPolicy(\Iset_\Player(\History^\timestep), \Action_\Player) \min \big(\overline \rho, \frac{\Policy(\Iset_\Player(\History^\timestep), \Action_\Player)}{\SamplingPolicy(\Iset_\Player(\History^\timestep), \Action_\Player)}\big)\\
    &\hspace*{25pt}\Big(\Reward_\Player^\timestep + \DiscountFactor \sum_{\History' \in \Histories} p(\History' | \History^\timestep, \Action_\Player) \VtraceVal^{\Policy_{\overline \rho}}_\Player(\History') - \VtraceVal^{\Policy_{\overline \rho}}_\Player(\History^\timestep) \Big)  \\
     =& \sum_{\Action_\Player \in \Actions_\Player(\History^\timestep)} \Policy_{\overline \rho}(\Iset_\Player(\History^\timestep), \Action_\Player) \\&\hspace*{25pt}\Big(\Reward_\Player^\timestep + \DiscountFactor \sum_{\History' \in \Histories} p(\History' | \History^\timestep, \Action_\Player) \VtraceVal^{\Policy_{\overline \rho}}_\Player(\History') - \VtraceVal^{\Policy_{\overline \rho}}_\Player(\History^\timestep) \Big) \\
     &\hspace*{25pt}\sum_{\Action_\Player ' \in \Actions_\Player(\History^\timestep)} \min \big(\overline \rho \SamplingPolicy(\Iset_\Player(\History^\timestep), \Action_\Player ') , \Policy(\Iset_\Player(\History^\timestep), \Action_\Player ')\big) = 0
\end{align*}
The first part of the last equation is Bellman equation for $\VtraceOperator\VtraceVal^{\Policy_{\overline \rho}}_\Player$. Therefore we deduce $\VtraceOperator \VtraceVal^{\Policy_{\overline \rho}}_\Player = \VtraceVal^{\Policy_{\overline \rho}}_\Player$. \qedsymbol
\section{Game Rules} \label{app:gamerules}
\subsection{Leduc hold'em} \label{app:leducrules}
Leduc hold'em is a simplified version of a Texas hold'em poker. The game is played with a deck comprising six cards: two pairs of King, Queen, and Jack. The game starts with each player putting a single chip into the pot. 

The game consists of two rounds, where players take turns playing. Each turn, the player can either call, raise, or fold. When either player plays fold action, the game terminates, and the winner is the other player. By playing the call action, the player puts into pot chips so that the total amount of chips by both players in the pot is the same. This means that when the amount of chips in the pot is already the same, the player does not put any chips into it. Each round, the players can raise at most two times. In the first round, when either player plays raise action, it calls the other player and then puts two additional chips into the pot. In the second round, the additional chips increase to four. The round ends when both players play call action immediately after each other.

After the first round, the public card is revealed. After the second round, the game ends, and the player who holds the same card as the public card wins. Otherwise, the player with a higher rank card wins. The descending order of ranks is King, Queen, and Jack. The utility given to the winning player is the amount of chips the opponent puts into the pot.

\subsection{Imperfect Information Goofspiel} \label{app:goofspielrules}
Imperfect Information Goofspiel is a strategic card game where players are dealt a private hand of thirteen cards, each with a value ranging from one to thirteen. A deck containing the same set of thirteen unique point cards is placed face-up on the table.

During each round, players make secret bids by selecting one card from their hand. The player with the highest bid wins the round and claims the corresponding point card from the table. However, if multiple players happen to make the same highest bid, the point card is discarded, leaving all participants empty-handed for that round. Throughout the game, bids are \textbf{never} revealed to the other players.

The game progresses through thirteen rounds, after which all players have played all their cards, and there are no remaining point cards on the table. At this point, the game concludes, and the player with the highest sum of won point card values emerges as the winner.

In this work, we use two different versions of Goofspiel based on the shuffled order of the point cards. The first variation shuffles the point cards randomly. The second variation arranges the point cards from the highest value to the lowest, eliminating the chance player from the game, making the game much smaller. Original Goofspiel is played with thirteen cards. However, in this work, we also use Goofspiel, played with only five and eight private and point cards.
\subsection{Battleships} \label{app:battleshiprules}
Battleships is a game played by two players on two square grid boards of size ten. Each player sees only its board but does not see the opponent's board. The game starts with each player secretly placing five ships on its board. The ships have the following sizes: 5, 4, 3, 3, 2. 

After both players place their ships, the game proceeds with players taking turns to shoot at the opponent's board. These shootings are public, and the shooting player knows if the ship was missed, hit, or sunken. The ship is sunken if all the tiles that it occupies are hit. The player that sinks all the opponent's ships wins.

We have used different versions of the Battleships, played on smaller boards, and each player has fewer ships.

\section{Additional experimental results} \label{s:addexperiments} 
\subsection{Counterfactual values in gadget games} \label{s:addcfv}
This experiment aims to show that using the counterfactual values computed from history critic does not significantly worsen the solution compared to using them from previous resolves. We trained ten critic networks with the same hyperparameter setting for Leduc hold'em and Goofspiel with five cards and both descending point card ordering. Then, we solved each subgame separately, either by computing the counterfactual values for the gadget game from the history critic or reusing them from previous searches. In Leduc hold'em, we split the game before dealing a public card and after. In Goofspiel, each subgame had a depth limit of 1, so the value function is called after a single action by each player from the root of the subgame. The action in the gadget game is not counted towards the depth limit. We then combine these subgame policies into a policy for a whole game and compute the exploitability. 

\begin{figure}[h]
  \centering
  \begin{subfigure}{.235\textwidth}
      \centering
      \includegraphics[width=.99\linewidth]{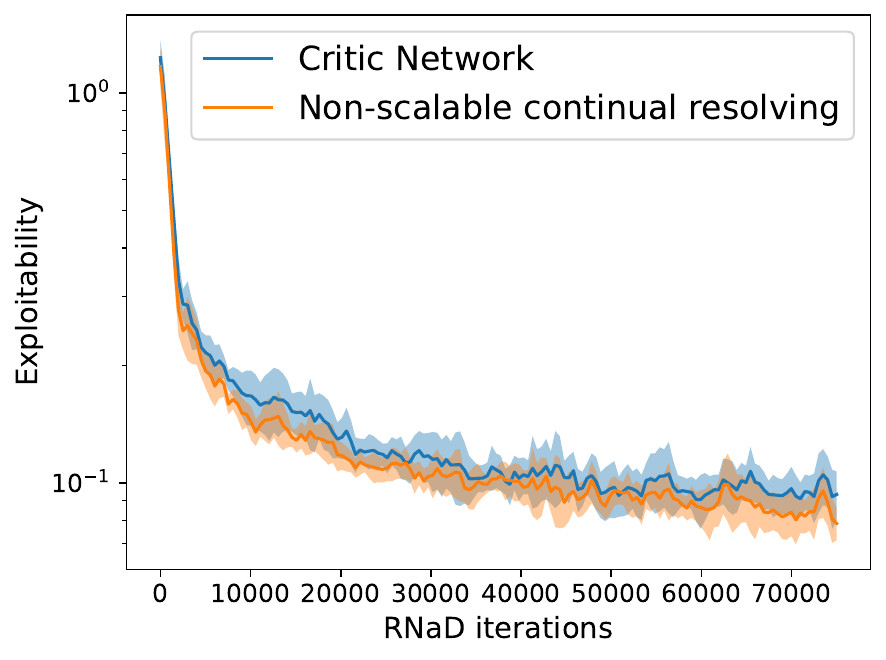}
      \caption{Leduc hold'em}
      \label{fig:leduccfv}
  \end{subfigure}
  \begin{subfigure}{.235\textwidth}
      \centering
      \includegraphics[width=.99\linewidth]{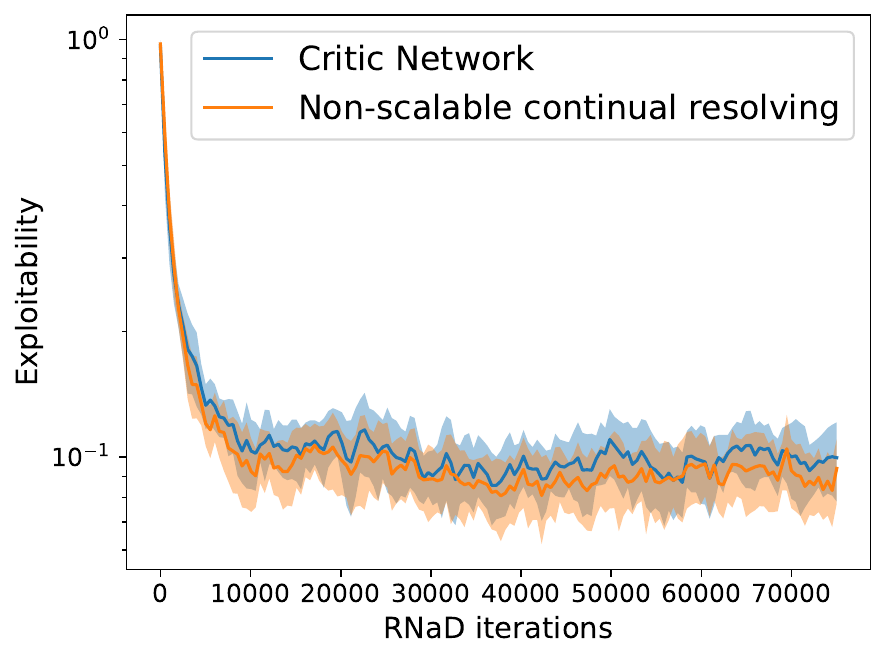}
      \caption{Goofspiel 5 descending}
      \label{fig:goofspielcfv}
  \end{subfigure}
  % \hspace{100pt}
  % \begin{subfigure}{.235\textwidth}
  %     \centering
  %     \includegraphics[width=.99\linewidth]{figures/exploitability_continual_leduc_P0.pdf}
  %     \caption{Goofspiel 5 randomized}
  %     \label{fig:leduccfv}
  % \end{subfigure}
  % \begin{subfigure}{.235\textwidth}
  %     \centering
  %     \includegraphics[width=.99\linewidth]{figures/exploitability_continual_battleship_2x2_2_6_P0.pdf}
  %     \caption{Battleships}
  %     \label{fig:goofspielcfv}
  % \end{subfigure}
  \caption{Exploitability based on the training iterations, when using the counterfactual values from previous searches or computing them from history critic.}
  \label{fig:cfvexperiment}
  % \Description{Goofspiel and Leduc Exploitability}
\end{figure}

The results are shown in \Cref{fig:cfvexperiment}, where we show both approaches' mean and the 95\% confidence interval. Using the counterfactual values from previous searches only slightly outperforms using the history critic. The results confirm that using critic network affects the quality of the solution in a minor way. With this insight, the remainder of our experiments were conducted with the computation of the counterfactual values from the critic network.

\subsection{Predefined transformations}
\begin{figure}[h]
  \centering
  \begin{subfigure}{.235\textwidth}
      \centering
      \includegraphics[width=.99\linewidth]{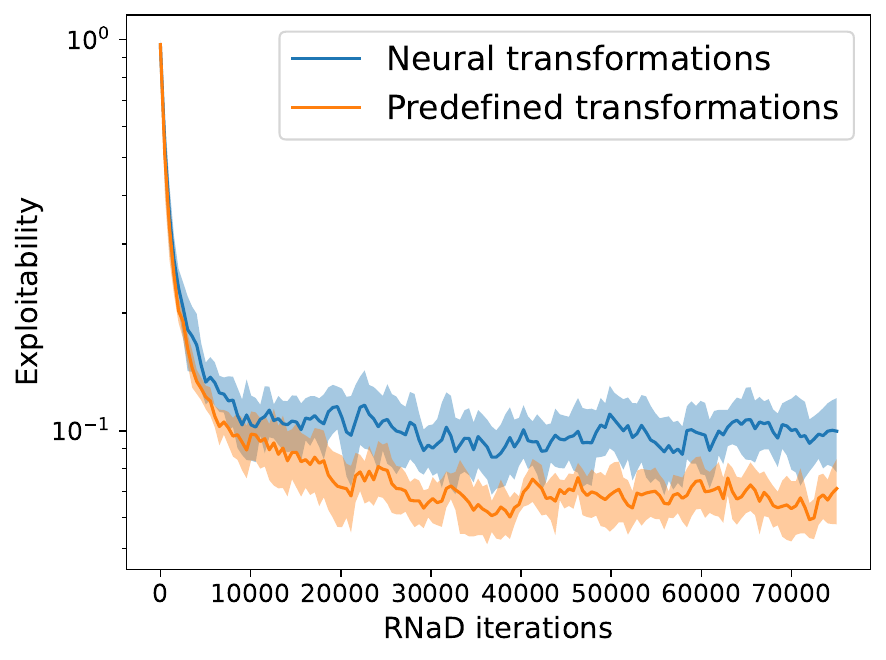}
      \caption{Goofspiel 5 descending}
      \label{fig:predefineddescending}
  \end{subfigure}
  \begin{subfigure}{.235\textwidth}
      \centering
      \includegraphics[width=.99\linewidth]{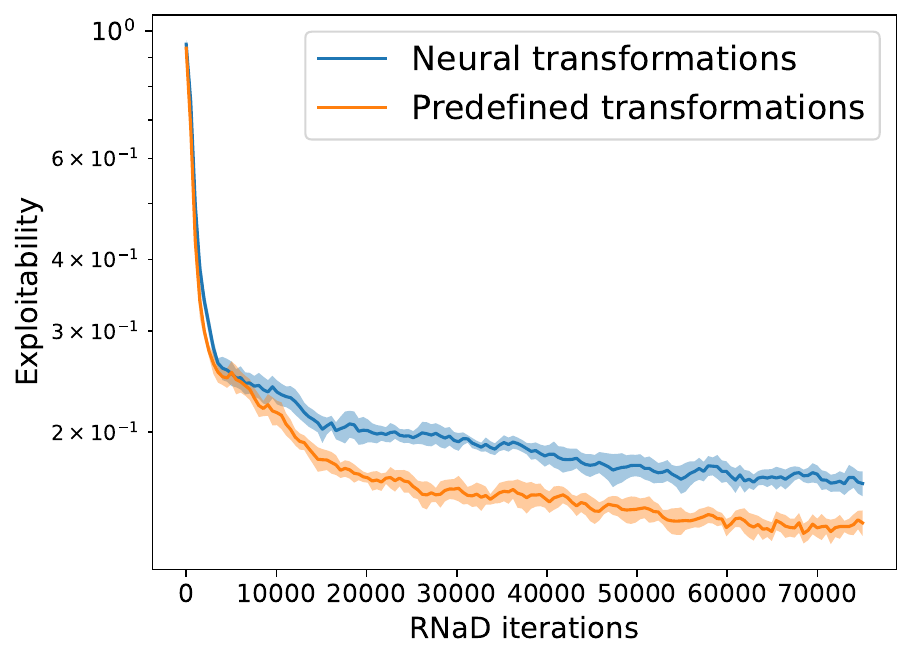}
      \caption{Goofspiel 5 randomized}
      \label{fig:predefinedrandom}
  \end{subfigure}
  % \hspace{100pt}
  % \begin{subfigure}{.235\textwidth}
  %     \centering
  %     \includegraphics[width=.99\linewidth]{figures/exploitability_continual_leduc_P0.pdf}
  %     \caption{Goofspiel 5 randomized}
  %     \label{fig:leduccfv}
  % \end{subfigure}
  % \begin{subfigure}{.235\textwidth}
  %     \centering
  %     \includegraphics[width=.99\linewidth]{figures/exploitability_continual_battleship_2x2_2_6_P0.pdf}
  %     \caption{Battleships}
  %     \label{fig:goofspielcfv}
  % \end{subfigure}
  \caption{Exploitability in Goofspiel based on the training iterations, comparing predefined transformations with the neural network transformations}
  \label{fig:predefinedexperiment}
  % \Description{Goofspiel and Leduc Exploitability}
\end{figure}

We have created 22 predefined transformations for Goofspiel. These transformations correspond to some strategies in Goofspiel that are strong in certain scenarios, like playing more higher value cards. We have then trained ten networks with the same trajectories. For each 500 iterations, we compared the exploitability when having a search depth-limit one between these predefined transformations and the neural network transformations. The results for Goofspiel 5 with descending and randomized order of point cards are shown in \Cref{fig:predefinedexperiment}.

The predefined transformations perform better in the tested scenario. This is to be expected, since the predefined transformations were specifically selected to perform well for Goofspiel. Therefore, \AlgorithmShort{} performance could be further improved by specyfing domain-specific transformations, or by improving the automatic transformations.

\section{Additional experimental details} \label{app:details}
We used the same setting for all the hyperparameters for all the experiments. The setting is shown in \Cref{tab:hyperparameters}. For the majority of hyperparameters, we adopted the settings established by the original authors in their work on RNaD \cite{perolat2022mastering}. The differences are setting importance sampling clipping $\overline \rho$ to $\infty$ to ensure that the estimate of V-trace is unbiased. Furthermore, we changed the rate of updating the regularization policy, learning rate, batch size, and size of hidden layers to compensate for training on smaller games than the original authors did with Stratego. As a sampling policy $\SamplingPolicy$, we have used the actor policy, effectively avoiding importance sampling clipping by working on-policy. 

The only exception with different hyperparameter settings was Battleship 7x7, for which we have set the importance sampling clipping $\overline \rho = 3$.
\begin{table}[!ht]
    \centering
    \begin{tabular}{ll}
        Parameter & Value \\ \hline
        reward regularization constant $\eta$ & 0.2\\
        regularization policy change $\Delta_m$ & 500 for first 200 steps \\
            & 10000 afterwards\\
        learning rate $\alpha_l$ & 0.0003\\
        gradient clip $c_g$ & 10000 \\
        NeuRD threshold $\beta$ & 2 \\
        NeuRD clipping $c_n$ & 10000 \\
        importance sampling clipping $\overline c$ & 1 \\
        importance sampling clipping $\overline \rho$ & $\infty$ \\
        decay rate $b_1$ in ADAM & 0 \\
        decay rate $b_2$ in ADAM & 0.999\\
        $\epsilon$ in ADAM & $10^{-8}$\\
        target network update $\gamma_a$ & 0.001\\
        batch size & 64 \\
        hidden layers size & 1024, 1024\\
        CFR+ iterations in resolving & 800\\
        Policy transformations & 10
    \end{tabular}
    \caption{Parameters used during evaluation of SePoT}
    \label{tab:hyperparameters}
\end{table}

In \Cref{tab:domainsize}, we provide approximate amount of unique histories for each of the used games during our experiments
\begin{table}[!ht]
    \centering
    \begin{tabular}{ll}
        Game & Histories \\ \hline
        Goofspiel 5 descending points  & $1.8 \cdot 10^4$ \\
        Goofspiel 5 randomized points  & $2.4 \cdot 10^6$ \\
        % Goofspiel 8 descending points  & $2 \cdot 10^9$ \\
        Goofspiel 8 randomized points  & $9.1 \cdot 10^{13}$ \\
        Goofspiel 13 descending points  & $4.9 \cdot 10^{19}$ \\
        Goofspiel 13 randomized points  & $3.4 \cdot 10^{28}$ \\
        Leduc hold'em  & $9.4 \cdot 10^3$ \\
        % Battleship 2x2  & $6.8 \cdot 10^{3}$ \\
        Battleship 3x3  & $ \approx 10^{14}$ \\
        Battleship 5x5  & $ \approx 10^{57}$ \\
        Battleship 7x7  & $ \approx 10^{138}$ 
    \end{tabular}
    \caption{Approximate amount of histories in tested games}
    \label{tab:domainsize}
\end{table}

\section{Pseudocodes} \label{app:pseudocodes}
To improve the understandability of the full algorithm, we provide pseudocodes for training \AlgorithmShort{} with some arbitrary trajectory-sampling-based policy-gradient algorithm and for gameplaying with CFR+. It is important to note that within CFR+, which performs a search, the counterfactual values have to be computed via the value function after the depth limit. This consists of retrieving multi-valued states for each history in the leaves and then computing the opponents' best response value for each infosets of each player.
\begin{algorithm}[h]
\caption{SePoT training}\label{alg:training}
\begin{algorithmic}[1]
\Input Sampling Policy $\PseudocodeSamplingPolicy$, Iterations $\PseudocodeIterations$
% \hspace*{\algorithmicindent} \textbf{Output} \\
\State $\PseudocodePolicyWeights, \PseudocodeTransformationWeights, \PseudocodeCriticWeights \gets$ InitializeNetworks()
\For{$t \gets 1 \text{ to } \PseudocodeIterations$}
    \State $\PseudocodeTrajectory \gets$ SampleTrajectories($\PseudocodeSamplingPolicy$)
    \State $\PseudocodePolicy_{\text{old}} \gets$ RetrievePolicyFromNetwork($\PseudocodeTrajectory, \PseudocodePolicyWeights$)
    \State $\PseudocodePolicyWeights \gets$ PolicyGradientUpdate($\PseudocodeTrajectory, \PseudocodeSamplingPolicy, \PseudocodePolicyWeights$) 
    \State $\PseudocodePolicy_{\text{new}} \gets$ RetrievePolicyFromNetwork($\PseudocodeTrajectory, \PseudocodePolicyWeights$)
    \State $\PseudocodePolicyDirection \gets \frac{\PseudocodePolicy_{\text{new}} - \PseudocodePolicy_{\text{old}}}{||\PseudocodePolicy_{\text{new}} - \PseudocodePolicy_{\text{old}}||}$
    \State $\PseudocodeTransformationWeights \gets$ UpdateTransformationNetwork($\PseudocodeTrajectory, \PseudocodePolicyDirection, \PseudocodeTransformationWeights$) 
    \State $\PseudocodePolicyTransformed \gets$ ApplyTransformations($\PseudocodePolicy, \PseudocodeTransformationWeights$ )
    \State $\PseudocodeCriticWeights \gets$ UpdateCriticNetwork($\PseudocodeTrajectory, \PseudocodeSamplingPolicy, \PseudocodePolicyTransformed, \PseudocodeCriticWeights$)
\EndFor
\State \Return$\PseudocodePolicyWeights,  \PseudocodeCriticWeights $
\end{algorithmic}
\end{algorithm}

\begin{algorithm}[h]
\caption{SePoT gameplay}\label{alg:gameplay}
\begin{algorithmic}[1]
\Input Actor Weights $\PseudocodePolicyWeights$, Critic Weights $\PseudocodeCriticWeights$, Player $\PseudocodePlayer$ Current decision node $\PseudocodeIset_\PseudocodePlayer$
% \hspace*{\algorithmicindent} \textbf{Output} \\
\If {CanPublicStateBeReconstructed($\PseudocodeIset_\PseudocodePlayer$)}
    \State $\PseudocodeIset_\PublicIndex \gets$ ReconstructPublicState($\PseudocodeIset_\PseudocodePlayer$)
    \State $\PseudocodeRange_\PseudocodePlayer, \PseudocodeRange_\ChancePlayer \gets$ RetrieveRanges($\PseudocodeIset_\PublicIndex, \PseudocodePolicyWeights$)
    \State $\PseudocodeCriticValues \gets$ RetrieveMultiValuedStates($\PseudocodeIset_\PublicIndex, \PseudocodeCriticWeights$)
    \State $\PseudocodeCounterfactualValues_{\Opponent} \gets$ ComputeOpponentsRootCFVs($\PseudocodeCriticValues, \PseudocodeRange_\PseudocodePlayer, \PseudocodeRange_\ChancePlayer$)
    \State $\PseudocodeGadgetGame \gets$ ConstructGadgetGame($\PseudocodeRange_\PseudocodePlayer, \PseudocodeRange_\ChancePlayer, \PseudocodeCounterfactualValues_{\Opponent} $)
    \State $\PseudocodePolicy \gets$ CounterfactualRegretMinimizationPlus($\PseudocodeGadgetGame, \PseudocodeCriticWeights$)
\Else
    \State $\PseudocodePolicy \gets$ RetrievePolicyFromNetwork($\PseudocodeIset, \PseudocodePolicyWeights$)
\EndIf
\State \Return$\PseudocodePolicy$
\end{algorithmic}
\end{algorithm}
\section{\AlgorithmShort{} components diagram}
\begin{figure}[H]
    \centering
    \includegraphics[width=\linewidth]{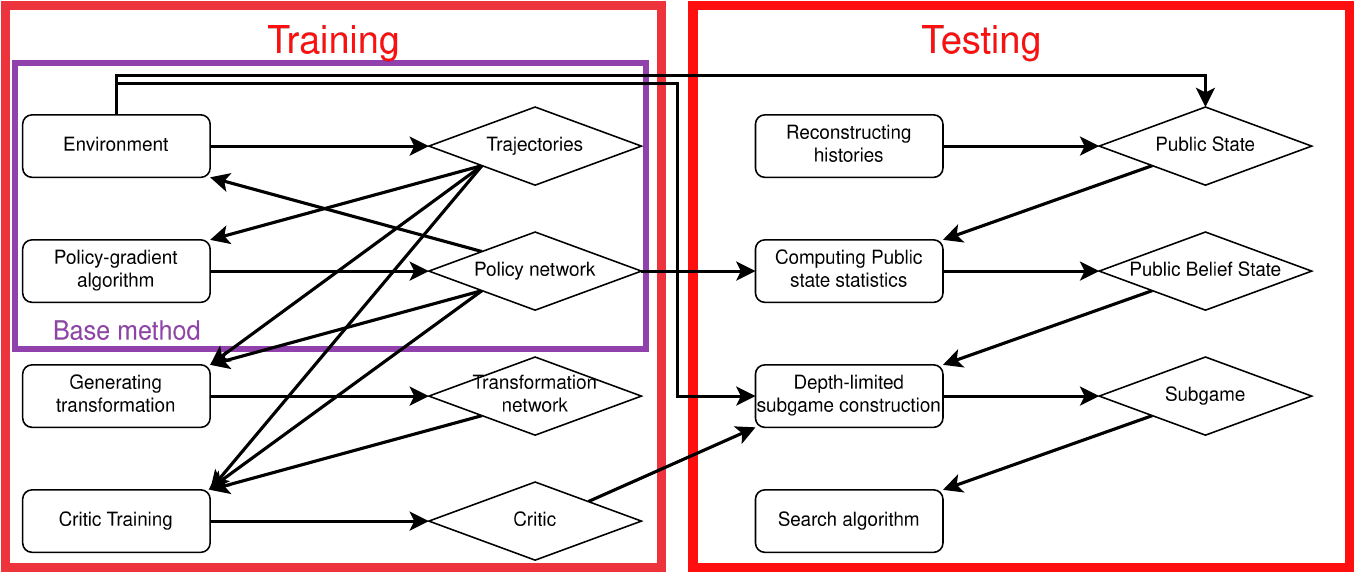}
    \caption{Main components of \AlgorithmShort{}}
    \label{fig:sepot_diagram}
\end{figure}

\AlgorithmShort{} consists of multiple components that interact together. In \Cref{fig:sepot_diagram} we show these components and their connection. The process of the SePoT can be summarized as follows: During training, the trajectories are sampled from the environment. These are then used to train the policy network with the underlying policy gradient algorithm. We generate the transformations for the same trajectories based on the policy network and train the transformation network with them. All three of policy and transformations from networks and trajectories are then used to train the estimate the expected value for each transformation, which is used for training the critic. This process is repeated for a specified amount of training iterations. During gameplay, when a player is about to make a decision, it either decides to sample action based on the policy network or refine the strategy with search. When refining the strategy, the algorithm first reconstructs the public state, which is all the histories that share the same public knowledge. Then, it uses a policy network to compute the reach of each infoset. After that, the depth-limited subgame is constructed, which requires computing counterfactual values at the root and giving the opponent a choice between different strategies after the depth limit. For both of these, we use the trained critic. This depth-limited subgame is then solved with some search algorithm; in our case, it was CFR+.

